\documentstyle[epsfig,12pt]{article}  
\newcommand{\beq}{\begin{equation}}
\newcommand{\eeq}{\end{equation}}
\newcommand{\bea}{\vspace{0.25cm}\begin{eqnarray}}
\newcommand{\eea}{\end{eqnarray}}

\newcommand{\r}{\mbox{{\boldmath
$\rho$}}}

\newcommand{\qb}{\mbox{{\bf
q}}}
\newcommand{\pb}{\mbox{{\bf
p}}}

\newcommand{\kb}{\mbox{{\bf
k}}}

\def\lsim{\mathrel{\rlap{\lower4pt\hbox{\hskip1pt$\sim$}}
    \raise1pt\hbox{$<$}}}         
\def\gsim{\mathrel{\rlap{\lower4pt\hbox{\hskip1pt$\sim$}}
    \raise1pt\hbox{$>$}}}         
\begin{document}
\vspace*{-2cm}
 
\bigskip

\begin{center}

\renewcommand{\thefootnote}{\fnsymbol{footnote}}

  {\Large\bf
Parton energy loss 
in an expanding quark-gluon plasma: 
Radiative vs collisional
\\
\vspace{.7cm}
  }
\renewcommand{\thefootnote}{\arabic{footnote}}
\medskip
{\large
  B.G.~Zakharov
  \bigskip
  \\
  }
{\it
 L.D.~Landau Institute for Theoretical Physics,
        GSP-1, 117940,\\ Kosygina Str. 2, 117334 Moscow, Russia
\vspace{1.7cm}\\}

  {\bf
  Abstract}
\end{center}
{
\baselineskip=9pt
We perform a comparison of the radiative and collisional parton 
energy losses in an expanding quark-gluon plasma.
The radiative energy loss is calculated 
within the light-cone path integral approach \cite{Z1}. The collisional
energy loss is calculated using the Bjorken method with an accurate
treatment of the binary collision kinematics.
Our numerical results demonstrate that for RHIC and LHC conditions
the collisional energy loss is relatively small in comparison to the
radiative one. We find an enhancement of the  
heavy quark radiative energy loss 
as compared to that of the light quarks
at high energies.
\vspace{.5cm}
\\
}

\noindent{\bf 1.} 
The suppression of high-$p_{T}$ hadrons in $AA$-collisions 
(usually called jet quenching)
observed in the experiments at RHIC (for a review, 
see \cite{RHIC_data})
is widely believed to be due to the parton energy loss in the hot quark-gluon
plasma (QGP) produced at the initial stage of nucleus-nucleus collision. 
The parton energy loss may come from the collisional energy loss 
and the induced gluon radiation.
The first estimate of the collisional energy loss in the QGP has been
done by Bjorken \cite{Bjorken1}. The radiative energy loss 
has been under active investigation 
in the last years  \cite{BDMPS,Z1,Z2,Z_YAF,Z3,GLV1,W1} 
(for a review, see \cite{BSZ}).
The calculations of the radiative energy loss within 
the BDMPS \cite{BDMPS,BSZ} and  the light-cone path integral (LCPI) 
\cite{Z1,Z_YAF,Z3}) approaches demonstrate that for high energy 
partons the energy loss is likely dominated by the induced gluon radiation.
The estimates given in \cite{Z04_RAA} show that 
the collisional energy loss can roughly increase the energy
loss by 30-40\% for RHIC energies.
However, it has recently been claimed \cite{Mustafa-Thoma,Mustafa,Abhee}
that for RHIC conditions the collisional energy loss may be as important as 
the radiative one, or even dominate 
at low energies. But an accurate comparison of the two mechanisms of the
energy losses so far has not been 
performed, say, in \cite{Abhee} even for the plasma 
with a uniform density the radiative energy loss was calculated incorrectly.
The authors have used the kinematical suppression factor for the radiative
energy loss obtained in \cite{GLV1} which strongly overestimates the kinematic 
suppression \cite{Z_kin}.
The question of the importance of the collisional energy loss is becoming of 
special current interest in connection with the recent data on the 
non-photonic electrons \cite{charm} indicating that 
the nuclear suppression for heavy quarks may be similar to that 
for light ones. This fact seems to be inconsistent
with purely radiative energy loss 
which, at low energies, should be suppressed 
for heavy quarks by the mass effects  \cite{Z_Moriond98}.  
Unfortunately, presently 
for both the radiative and collisional losses the uncertainties in the 
theoretical predictions are rather big. Say, the results are very 
sensitive to the choice of $\alpha_{s}$ 
(running or constant \cite{Z04_RAA,Peshier}), to the infrared effects 
\cite{Z04_RAA}.
For clarifying the situation with the relative contributions of the 
radiative and collisional energy loss it is important 
to perform the calculations with the same 
parametrizations of the coupling constant and 
the infrared cutoffs. 
In the present paper we present the results of such calculations for expanding
QGP for RHIC and LHC conditions.

To evaluate the radiative energy loss we use the LCPI formalism
\cite{Z1,Z_YAF,Z3}.
It accurately treats  the mass and finite-size effects, and applies
at arbitrary strength of the Landau-Pomeranchuk-Migdal
(LPM) effect \cite{LPM}. Other available approaches
have limited domains of applicability, and can only be used either in 
the regime of strong (the BDMPS formalism \cite{BDMPS})
or weak (the GLV formalism \cite{GLV1}) LPM suppression (the GLV 
approach \cite{GLV1}, in addition, is restricted to the emission of soft
gluons). 

Following Bjorken 
\cite{Bjorken1} we evaluate the collisional energy loss
for elastic binary collisions. However, contrary to the Bjorken analysis
we treat accurately the kinematics of the binary collisions and 
fluctuations of the momentum transfer due to the thermal parton motion
in the QGP. For the infrared momentum cutoff we use the Debye mass
included into the gluon propagator. In pQCD more accurate 
treatment of the infrared region of collective excitations is possible in 
the Hard Thermal Loop (HTL) resummation technique \cite{HTL}. 
However, it is unlikely that the pQCD HTL formalism assuming that 
$g\ll 1$ is reliable
for RHIC and LHC conditions when the plasma temperature 
$T\lsim (2-3)T_{c}$ ($T_{c}\approx 170$ MeV is the
temperature of the deconfinement phase transition) and $g\sim 1.5-2$.
Say, in the leading order in $g$ the HTL approximation predicts 
zero magnetic screening mass. However, the lattice calculations 
\cite{magnetic_Md} 
show that the magnetic mass may be of the order of 
the electric Debye mass.
Also, the HTL approach to  
the collisional energy loss \cite{DE-HTL,DE-Thoma} does not 
incorporate the running $\alpha_{s}$ which turns out to be very 
important \cite{Peshier}. 

In our calculations of the radiative and collisional energy loss
the infrared cutoffs for the $t$-channel gluon exchanges 
are performed in the same way. 
This should minimize the theoretical uncertainties associated  with
the collective excitations in the ratio of 
the radiative to collisional energy loss which is of interest in
the present paper.
Numerical calculations are performed with the Debye mass obtained in the 
lattice calculations.
Our results show that for RHIC and LHC conditions the effect of
the collisional energy loss is relatively small, and cannot be crucial for
the difference in the nuclear suppression of the heavy and light quark jets.
For $c$-quark we find the radiative energy loss which 
is very close to that for the light quarks. For $b$-quark
the radiative energy loss is significantly suppressed at
$E\lsim 20$ GeV, but at high energies it exceeds the charm (and light quark)
energy loss. 
Our results show that the observed at RHIC suppression of the non-photonic 
electrons \cite{charm} may be naturally described if in the experimentally 
studied 
region $p_{T}\lsim 8$ GeV the non-photonic electrons are dominated by charm
production.
Although the pQCD calculations \cite{CNV} predict that the bottom contribution
dominates at $p_{T}\gsim 5$ GeV, this possibility cannot be excluded
since the pQCD results in the charm mass region are very fragile, 
say, the calculations of Ref. \cite{CNV} underestimates the normalization of 
the experimental electron spectrum by a factor of about 5 \cite{Xu}. 
The same situations
occurs for the $D$-meson production \cite{Tai}. Thus, presently, it is 
not clear at all whether the pQCD calculations predict correctly 
the ratio charm/bottom for the kinematical region studied at RHIC.
  
\vspace{.2cm}
\noindent{\bf 2.} 
We begin with the radiative energy loss. For definiteness we consider 
the case of an energetic quark. We assume that the fast quark, $Q$, 
produced in a hard process at $z=0$, passes through a length $L$ of an 
expanding QGP (we choose the $z$-axis
along the initial quark momentum). We define the energy loss as
\beq
\Delta E=E\int\limits_{x_{min}}^{x_{max}} dx x\frac{dP}{dx}\,,
\label{eq:10}
\eeq
where $E$ is the initial quark energy, $x$ is the Feynman
variable for the radiated gluon, and $dP/dx$ is the probability distribution
of the induced gluon emission. Since for 
hard gluons with $x\gsim 0.5$ the jet really does not disappear, 
from the point of view of the jet quenching, a reasonable choice for the upper
limit of $x$-integration is $x_{max}=0.5$. 
For $x_{min}$ we use the value
$m_{g}/E$, where hereafter $m_{g}$ is the gluon quasiparticle mass
(the quark quasiparticle mass will be denoted $m_{q}$). 

In the LCPI approach \cite{Z1} $dP/dx$  can be expressed in terms of
the Green's function for a two-dimensional Schr\"odinger
equation in the impact parameter space in which  the longitudinal
coordinate $z$ plays the role of time. This Schr\"odinger
equation describes evolution of the light-cone wave function of a
spurious three-body $q\bar{q}g$ color singlet system. 
The corresponding Hamiltonian for the medium number density $n(z)$ 
has the form
\beq
H=
-\frac{1}{2M(x)}\left(\frac{\partial}{\partial \r}\right)^{2}
-i\frac{n(z)\sigma_{3}(\rho,x)}{2}
+\frac{1}{L_{f}}\,.
\label{eq:20}
\eeq
Here
$
M(x)=Ex(1-x)\,
$
is the reduced "Schr\"odinger mass", $L_{f}=2M(x)/\epsilon^{2}$
with $\epsilon^{2}=m_{q}^{2}x^{2}+m_{g}^{2}(1-x)^{2}$,
$\sigma_{3}(\rho,x)$ is the cross section of interaction
of the $q\bar{q}g$ system with a medium constituent which reads 
$
\sigma_{3}(\rho,x)=\frac{9}{8}
[\sigma_{q\bar{q}}(\rho)+
\sigma_{q\bar{q}}((1-x)\rho)]-
\frac{1}{8}\sigma_{q\bar{q}}(x\rho)\,
$
\cite{NZZ,Z1},
where
\beq
\sigma_{q\bar{q}}(\rho)=\alpha_{s}^{2}C_{T}C_{F}\int d\qb_{\perp}
\frac{[1-\exp(i\qb_{\perp}\r)]}{(q^{2}_{\perp}+\mu^{2}_{D})^{2}}\,
\label{eq:30}
\eeq
is the dipole cross section for the color singlet $q\bar{q}$ pair
($C_{F,T}$ are the color Casimir for the quark and thermal parton 
(quark or gluon), $\mu_{D}$ is the Debye mass).
The gluon spectrum may be written as
\cite{Z_Moriond98,Z04_RAA} 
\beq
\frac{d P}{d
x}=
\int\limits_{0}^{L}\! d z\,
n(z)
\frac{d
\sigma_{eff}^{BH}(x,z)}{dx}\,,
\label{eq:40}
\eeq
\beq
\frac{d
\sigma_{eff}^{BH}(x,z)}{dx}=\mbox{Re}
\int\limits_{0}^{z} dz_{1}\int\limits_{z}^{\infty}dz_{2}\int d\r\,
\hat{g}K_{v}(z_{2},\r_{2}|z,\r)
\sigma_{3}(\rho,x)
K(z,\r|z_{1},\r_{1}){\Big|}_{\r_{1}=\r_{2}=0}\,.
\label{eq:50}
\eeq
where $K$ is the Green's function for the Hamiltonian (\ref{eq:20}), and 
$K_{v}$ is the Green's function for the same Hamiltonian with $n=0$,
$
\hat{g}=
{\alpha_{s}
 C_{F}[1+(1-x)^{2}]}
{[2xM^{2}(x)]^{-1}}\,
\frac{\partial}{\partial \r_{1}}\cdot\frac{\partial}{\partial \r_{2}}\,
$ is the vertex operator.
The $d\sigma^{BH}_{eff}/dx$ can be viewed as an 
effective Bethe-Heitler
cross section, which accounts for the LPM and finite-size effects.
It can be represented as \cite{Z04_RAA}
\beq
\frac{d
\sigma_{eff}^{BH}(x,z)}{dx}=-\frac{\alpha_{s} C_{F}[1+(1-x)^{2}]}
{\pi x M(x)}\mbox{Im}
\int\limits_{0}^{z} d\xi
\left.\frac{\partial }{\partial \rho}
\left(\frac{F(\xi,\rho)}{\sqrt{\rho}}\right)
\right|_{\rho=0}\,\,,
\label{eq:60}
\eeq
where the function $F$ is the solution to the radial Schr\"odinger 
equation for the azimuthal quantum number $m=1$ (we omit the argument $x$)
\beq
i\frac{\partial F(\xi,\rho)}{\partial \xi}=
\left[-\frac{1}{2M}\left(\frac{\partial}{\partial \rho}\right)^{2}
-i\frac{n(z-\xi)\sigma_{3}(\rho)}{2}+
\frac{4m^{2}-1}{8M\rho^{2}}
+\frac{1}{L_{f}}
\right]F(\xi,\rho)\,.
\label{eq:70}
\eeq
The boundary condition for $F(\xi,\rho)$ reads
$F(\xi=0,\rho)=\sqrt{\rho}\sigma_{3}(\rho)
\epsilon K_{1}(\epsilon \rho)$, where 
$K_{1}$ is the Bessel function.
The time variable $\xi$ in (\ref{eq:70}), in terms of the variables $z$ 
and $z_{1}$
of equation (\ref{eq:50}), is given by $\xi=z-z_{1}$; i.e.,
contrary to the Schr\"odinger equation for the Green's functions
entering (\ref{eq:50}),  (\ref{eq:60}) represents the spectrum 
through the solution
to the Schr\"odinger equation, which describes evolution of the $q\bar{q}g$
system back in time. It allows one to have a smooth boundary condition,
which is convenient for numerical calculations.

The above equations are written for fixed $\mu_{D}$ and 
$\alpha_{s}$. For $z$-dependent $\mu_{D}$ the dipole cross section,
besides $\rho$, depends on $z$ as well. The incorporation of the $z$-dependence
of the dipole cross section  does not lead to any
problems. To account for the effect of running $\alpha_{s}$ on the dipole
cross section we include $\alpha_{s}(q^{2})$ in the integrand in 
(\ref{eq:30}).
The inclusion of the running $\alpha_{s}$ in the vertex operator 
$\hat{g}$
which corresponds to the emitted gluon is a more delicate question since
(\ref{eq:50}) is written in the coordinate representation, and there is no 
explicit dependence on the parton transverse momenta in the vertex
$q\rightarrow gq$. To generalize the formula (\ref{eq:50}) to the running
$\alpha_{s}$ we use the Schr\"odinger diffusion relation 
$\rho\sim \sqrt{(z-z_{1})/M(x)}$
connecting the longitudinal scale $z-z_{1}$ and $\rho$ scale in Eq. 
(\ref{eq:50}). 
In terms of the transverse separation between quark and gluon 
$\rho$ the transverse gluon momentum can be estimated via the uncertainty
relation $q\sim 1/\rho$.
Thus it seems quite reasonable to use for the virtuality scale in the 
splitting vertex the parametrization $q^{2}\approx a M(x)/(z-z_{1})$.
We adjusted the coefficient $a$ to reproduce in our formalism 
the $N=1$ rescattering contribution (which dominates
the gluon spectrum) evaluated in the diagrammatic approach in momentum
representation \cite{Z_kin}. For our parametrization of the running $\alpha_{s}$
(discussed in Sec. 4) it gives $a\approx 1.85$.

The induced gluon spectrum can also be used to estimate the effect of the 
energy gain due to absorption of the thermal gluons. For the plasma with fixed
($z$-independent) temperature in the collinear gluon 
approximation one can obtain for
the energy gain
\beq
\Delta E_{gain}\approx E\int\limits_{x_{min}}^{x_{max}} dx x(1+x) n_{B}(Ex)
\frac{dP}{dy}(y=x/(1+x),E')\,,
\label{eq:80}
\eeq
where $E'=E(1+x)$ is the parton energy after thermal gluon absorption, 
$n_{B}(p)=(e^{p/T}-1)^{-1}$ is the thermal gluon distribution. 
An accurate evaluation of this effect for an expanding plasma in the 
situation when the plasma density changes strongly at the gluon formation 
scale $L_{f}$
is a complicated problem. To estimate the effect of gluon absorption 
we have simply used for the thermal distribution the distribution averaged 
over $z$.
Our numerical calculations show that for RHIC and LHC conditions 
the effect of the gluon absorption is suppressed by about two orders of 
magnitude as compared to the gluon emission. 
For this reason it can be safely neglected, and the difficulties with 
its evaluation are not important for the jet quenching 
phenomenology.

\vspace{.2cm}
\noindent{\bf 3.}
Let us now discuss the collisional energy loss. As we said we calculate 
it through the energy transfer in the binary elastic collisions of the 
fast parton with the thermal quarks and gluons treated as free particles. 
In this approximation the collisional energy loss 
per unit length can be written as
\bea
\frac{dE}{dz} = \frac{1}{2Ev}\sum_{p=q,
g} g_{p} 
\, \int \frac{d\pb'}{2E'(2\pi)^{3}}
\int \frac{d\kb\, n_p (k) }{2k(2\pi)^{3}}\nonumber\\
\times
\int \frac{d\kb'[1+\epsilon_p n_p (k')] }{2k'(2\pi)^{3}}
(2\pi)^{4}\delta^{4}(P+K-P'-K')\omega 
\langle |M(s,t)|^{2}\rangle \theta(\omega_{max}-\omega)
\label{eq:90}
\eea
where $v\approx1$ is the velocity of the fast quark,
$P=(E,\pb)$ and $K=(k,\kb)$ are the momenta for incoming partons,
$P'=(E',\pb')$ and $K'=(k',\kb')$ are the momenta for outgoing partons,
$\omega=E-E'$ is the energy transfer, $M(s,t)$ is the matrix element
for $Qp\rightarrow Qp$ scattering ($s=(P+K)^{2}$, 
$t=(P-P')^{2}$ are the Mandelstam variables), 
$n_q(k)=(e^{k/T}+1)^{-1}$ and $n_{g}(k)=(e^{k/T}-1)^{-1}$ are
the Fermi and Bose distributions,
$\epsilon_{q}=-1$, $\epsilon_{g}=1$, 
$ g_q =4 N_c N_f$, $g_g = 2(N^{2}_{c} -1)$.
In (\ref{eq:90}) $\omega_{max}$ is the upper
limit of the energy loss.
Similarly to the radiative energy loss we take
$\omega_{max}=E/2$.
After integrating over the $\pb'$, azimuthal angle of the transverse 
momentum  $\kb_{\perp}$ and $k'_{z}$, (\ref{eq:90}) takes the form
\bea
\frac{dE}{dz} = \frac{1}{16E E'v(2\pi)^{4}}\sum_{p=q,
g}g_{p}
\int \frac{dk_{z}dk_{\perp}k_{\perp}\,n_p (k) }{k}\nonumber\\
\times
\int \frac{d\qb_{\perp}[1+\epsilon_p n_p (k')] }{k'}
\frac{\omega}{J} 
\langle |M(s,t)|^{2} \rangle\theta(\omega_{max} - \omega)\,,
\label{eq:100}
\eea
\beq
J=\left|\frac{\partial}{\partial q_{z}}(E'+k')\right|
=\left|\frac{k'_{z}}{k'}+\frac{q_{z}-P_{z}}
{\sqrt{m_{Q}^{2}+(P_{z}-q_{z})^{2}+\qb_{\perp}^{2}}}\right|\,,
\label{eq:110}
\eeq
where 
$\qb=\kb'-\kb$.
The longitudinal component $q_{z}$ is determined from the energy
conservation $E+k=E'+k'$. 
At small $|t|$  the amplitude is dominated by the 
$t$-channel gluon exchange which gives for the average squared matrix element
\beq
\langle |M(s,t)|^{2}\rangle \approx C_p \frac{2\pi \alpha_s^2(|t|)}
{(|t|+\mu^{2}_{D})^{2}} \,
\label{eq:120} 
\eeq
with $C_q = \frac{N_c^2 - 1}{2 N^2_c}$, $C_g = 1$.
The $\omega$ may be written as (if we take $v=1$)
\beq
\omega=\frac{-t-t k_{z}/E+2\kb_{\perp}\qb_{\perp}}{2(k-k_{z})}\,.
\label{eq:130}
\eeq
Note that in the Bjorken analysis \cite{Bjorken1} the last two terms in 
the numerator of (\ref{eq:130}) have been neglected. 
In this case neglecting the
statistical Pauli-blocking and Bose enhancement factors one
can represent (\ref{eq:90}) as
\beq
\frac{dE}{dz} \approx \frac{1}{2(2\pi)^{3}}\sum_{p=q,g} g_{p}\, \int \, 
d\kb \frac{n_p (k)}{k}
\int \limits_{0}^{|t|_{max}} dt|t|\frac{d\sigma}{dt}\,,
\label{eq:140}
\eeq
where 
$
|t|_{max} \approx 2(k-k_{z})\omega_{max}\,.
$ 
Eq. (\ref{eq:140}) is convenient for numerical calculations. 
However, at low energies
$E\lsim 10$ GeV, it is not accurate enough. For this reason we use the 
form (\ref{eq:100}).
In numerical calculations
we have used accurate formulas for the matrix elements \cite{Owens,Combridge}.
Note that for heavy quark at $E\lsim m_{q}^{2}/T$ (in this energy region
the heavy quark becomes nonrelativistic in the centre of mass system of
the binary collision)
the value of $\omega_{max}$ may be smaller than $E/2$ due to the kinematical 
limits. It suppresses the energy loss at low energies. 

\vspace{.2cm}
\noindent{\bf 4.} 
To apply  our formulas we need to specify 
the parametrization of $\alpha_{s}$ and mass parameters.
We parametrize $\alpha_{s}(Q^{2})$ by the one-loop expression
which is frozen at some value $\alpha_{s}^{fr}$ at $Q\le Q_{fr}$. 
Previously such a form  with $\alpha_{s}^{fr}\approx 0.7$ was used in 
the analyses of the low-$x$ structure functions within the 
dipole approach \cite{NZ,NZ_HERA,NZZ}.
A similar parametrization has been used in \cite{DKT} 
in the analysis of the heavy quark jets.
From the analysis of the heavy quark energy loss in vacuum the authors 
of Ref. \cite{DKT} obtained
\beq
\int_{\mbox{0}}^{\mbox{\small 2 GeV}}\!dQ\frac{\alpha_{s}(Q^{2})}{\pi}
\approx 0.36 \,\,
\mbox{GeV}\,.
\label{eq:150}
\eeq  
For our parametrization from (\ref{eq:150}) one can obtain
$\alpha_{s}(Q<Q_{fr})=\alpha_{s}^{fr}\approx 0.7$, and 
$Q_{fr}\approx 0.82$ GeV (for $\Lambda_{QCD}=0.3$ GeV) which agree
surprisingly well with
the parameters of Refs. \cite{NZ,NZ_HERA}.
In the vacuum the stopping of the growth of $\alpha_{s}$ 
at $Q\lsim Q_{fr}\sim 1$ GeV may be caused by
the nonperturbative effects \cite{DKT,DMW}. In the QGP
thermal partons can give additional suppression of $\alpha_{s}$
at low momenta \cite{Gendenshtein}.
Unfortunately, at present there is no robust 
information on $\alpha_s(Q^{2})$ in the QGP for gluons interacting with 
the energetic ($E\gg T$) partons which is necessary in our case.
Available pQCD calculations are performed in the static limit 
(see, for example, \cite{BPS,ESW,CH} and references therein). 
The running coupling constant obtained in \cite{ESW,CH} has a pole 
at $Q/\Lambda_{QCD}\sim 3$ at $T\sim 250$ MeV. Thus, in pQCD, even 
for the static case, the situation with $Q$-dependence of the 
in-medium $\alpha_{s}$ is unclear.
In the absence of robust analytical theoretical 
predictions for the in-medium $\alpha_{s}$ for fast partons 
it seems reasonable to estimate $\alpha_{s}^{fr}$
from the lattice results on the thermal $\alpha_{s}(T)$.
The lattice simulations \cite{alpha-lattice} give  
$\alpha_{s}(T)$ smoothly decreasing  from $\sim 0.5$ at 
$T\approx 175$ MeV to $\sim 0.35$ at $T\approx 400$ MeV. This behaviour 
of $\alpha_{s}$ in the QGP is also consistent with the analysis of the 
lattice data within the quasiparticle model \cite{LH}. One can expect that 
the thermal $\alpha_{s}(T)$ should be somewhat smaller than $\alpha_{s}^{fr}$.
For this reason it seems reasonable to use $\alpha_{s}^{fr}\sim0.5$ 
for RHIC and LHC conditions.

The collisional energy loss of light quarks and gluons is not sensitive to 
the quark and gluon quasiparticle masses. However, this is not the case
for the induced gluon radiation which is especially sensitive to $m_{g}$.
In the pQCD HTL
resummation \cite{HTL} $m_{q}=gT/\sqrt{3}$ and  $m_{g}=gT\sqrt{(1+N_{f}/2)/2}$.
Since the HTL pQCD formulas may be unreliable for 
RHIC and LHC conditions
it seems better to 
to use the results of the lattice simulations.
We use the quasiparticle masses
obtained in Ref. \cite{LH} from the 
analysis of the lattice data within the quasiparticle model. 
For the relevant range of the plasma temperature $T\sim (1-3)T_{c}$ 
 the analysis 
\cite{LH} gives $m_{q}\approx 0.3$ and $m_{g}\approx 0.4$ GeV. 

Besides the quasiparticle masses, we need to specify the
the Debye mass which enters the dipole cross
section and the amplitudes of the binary collisions.
We perform calculations for a fixed and $T$-dependent 
$\mu_{D}$. In the first case 
we use the Debye mass obtained
in \cite{LH} through the perturbative relation $\mu_{D}=\sqrt{2}m_{g}$ 
with $m_{g}$ extracted from the quasiparticle fit of the lattice data,
which gives approximately the $T$-independent value $\mu_{D}\approx 0.57$ GeV. 
For the $T$-dependent parametrization we take
the Debye mass obtained
in the lattice calculations for $N_{f}=2$ \cite{Bielefeld_Md} which 
give the ratio $\mu_{D}/T$ slowly decreasing with $T$  
($\mu_{D}/T\approx 3$ at $T\sim 1.5T_{c}$, 
$\mu_{D}/T\approx 2.4$ at $T\sim 4T_{c}$).

For evolution of the QGP produced in $AA$-collisions we use 
the Bjorken model \cite{Bjorken2} with the 
longitudinal expansion which gives the proper time dependence
of the plasma temperature $T^{3}\tau=T_{0}^{3}\tau_{0}$ ($T_{0}$ is the
initial plasma temperature).
For fast partons produced in the central rapidity region 
of $AA$-collisions our coordinate $z$ equals the 
proper time $\tau$. Thus, we have the number density 
$n(z)\propto 1/z$ for $z>\tau_{0}$.
In the mixed phase the fraction of the QGP was calculated according to
the $1/\tau$ dependence of the entropy density. Rescatterings in the hadron
phase giving a small contribution have been neglected.
As in our earlier analysis of the nuclear modification factor \cite{Z04_RAA} 
we assume that the QGP is in the thermal and chemical equilibrium
(we take $N_{f}=2.5$).
For RHIC we performed calculations for
 $T_{0}=297$ MeV and $\tau_{0}=0.5$ fm 
which correspond to the initial conditions in successful hydrodynamic 
description of $Au+Au$ collisions at 
$\sqrt{s}=200$ GeV \cite{HeinzKolb},
and agree with the total entropy extracted from
the charged particle multiplicity. For $Pb+Pb$ collisions at LHC
for $\sqrt{s}=5500$ GeV we use  $T_{0}=350$ MeV (with the same 
$\tau_{0}=0.5$ fm), which was obtained from the extrapolation of 
the RHIC data on the charged particle distribution to the LCH energy 
performed in \cite{Busza}.

\vspace{.2cm}
\noindent{\bf 5.}
We present the numerical results for $\alpha_{s}^{fr}=0.5$.   
In Figs.~1,~2 we plot the radiative 
and collisional energy losses for the light quark and gluon
for RHIC and LHC conditions. To illustrate the effect of the running
coupling constant we present the results for the case 
$\alpha_{s}(Q)=\alpha_{s}^{fr}$ as well. 
The  higher panels correspond to 
the $T$-independent $\mu_{D}=0.57$ GeV \cite{LH}, and the lower panels
to the $T$-dependent Debye mass from Ref. \cite{Bielefeld_Md}.
The calculations were performed for
$L=5$ fm which is the typical parton pathlength in
the QGP (and mixed) phase with life-time about $R_{A}\sim 6$ fm
for the central heavy ion collisions.
One sees that the fraction of the collisional energy loss
is relatively small. At $E\sim 10$ GeV the ratio 
$\Delta E_{col}/\Delta E_{rad}$ is about 0.3-0.4 for quarks and 0.2-0.3 for
gluons.  The smaller fraction of the collisional
energy loss for gluons results from the additional color 
factor $C_{A}/C_{F}=9/4$ 
for the radiative energy loss for gluons (which is absent for the collisional
energy loss). For this reason the effect
of the collisional energy loss on the nuclear modification factor
should be weaker in the kinematical regions where the high-$p_{T}$ 
hadron spectra are dominated by gluon jets. The fraction of the 
collisional energy loss drops as energy increases.
The $T$-dependent parametrization of the 
Debye mass gives somewhat smaller $\Delta E$. It is due to suppression of 
the rescatterings
in the initial high-temperature region with $z\lsim 1$ fm 
where $\mu_{D}$ may be about 1 GeV. However, one can see that 
the sensitivity of the results to the Debye mass is relatively weak.
The results shown in Figs.~1,~2 demonstrate clearly the importance of the 
running $\alpha_{s}$ for both the radiative and collisional energy 
loss. It leads to flattening the energy losses at high energies. 

In Fig.~3 we show the results for the charm ($m_{c}=1.2$ GeV) and 
bottom ($m_{b}=4.5$ GeV) quarks obtained
with running $\alpha_{s}$ for the Debye mass from \cite{Bielefeld_Md}. 
The difference in the radiative 
energy loss for light and charm quarks is small.
The charm radiative energy loss is only suppressed by $\sim$ 10\% at 
$E\sim 10$ GeV compared to the light quarks. 
For the bottom quark the mass suppression at low energies 
is significant. Note that at high energies the radiative 
energy loss for the bottom
quark becomes larger than that for the charm quark, and the charm
contribution exceeds slightly the light quark one.  
This fact is connected with emission of gluons at moderate values of $x$
where the induced radiation for heavy quarks turns out to be enhanced
in the so-called diffusion regime when $L\ll L_{f}$ \cite{Z_OA}.    
A detailed discussion of this effect reflecting a complicated
interplay of the finite-size and mass effects in the induced gluon
emission will be given elsewhere. 

A small difference in the energy loss for the light and charm quarks show
that one can expect approximately
the same nuclear suppression for light and $c$-quarks.
This is in contradiction with the considerable suppression
of the induced gluon radiation from charm predicted in \cite{DH}. 
However, in \cite{DH} 
there was not performed any accurate evaluation
of the mass effects in the induced radiation. 
To obtain the heavy quark spectrum the authors multiplied the 
BDMPS spectrum obtained for massless partons in the oscillator approximation
by the suppression factor defined as the ratio of the vacuum gluon emission
spectra for heavy and light quarks which evidently has noting to do with
the mass modification of the induced gluon radiation. The enhancement
of the energy loss for heavy quarks at high energies is absent in \cite{DH}.

To study the infrared sensitivity of our results we also performed
computations for $m_{g}=0.2$ and $m_{g}=0.6$ GeV.
These values of $m_{g}$ give reasonable 
the lower and upper limits of the infrared cutoff for the induced gluon
emission for RHIC and LHC conditions\footnote{
Note that the analysis of the 
low-$x$ proton structure function within the dipole BFKL equation
\cite{NZZ,NZ_HERA} gives the value of the effective gluon mass for 
gluon emission in the parton-nucleon interaction about 0.75 GeV.
This value agrees well with the natural infrared cutoff 
for gluon emission in the vacuum $m_{g}\sim 1/R_{c}$, where 
$R_{c}\approx 0.27$ fm is the gluon correlation radius in the QCD vacuum
\cite{Shuryak1}. One can expect that in $AA$-collisions
the infrared cutoff will be approximately the same
only for gluon emission in
the developed mixed phase and for fast gluons with $L_{f}\gg L$ which
give relatively small contribution to the total radiative energy loss.}.
At $E\sim 10$ GeV for $m_{g}=0.2$ GeV $\Delta E_{rad}$ is bigger
by $\sim 20-30$\% than that for $m_{g}=0.4$ GeV, for $m_{g}=0.6$ GeV
the effect is of opposite sign. The effect of variation of the
gluon mass in the above interval of $m_{g}$ becomes small at higher energies
($\lsim 10-15$\% at $E\gsim 40$ GeV). 

To study the effect of variation of $\alpha_{s}$ we have also performed 
calculations using for $\alpha_{s}^{fr}$ the values 0.7
and 0.35. The first one  
neglects the in-medium suppression of coupling constant, and the second
one, in the light of the lattice results, can be viewed as an low bound for 
$\alpha_{s}^{fr}$ for RHIC and LHC conditions. Numerically we obtained 
approximately 
the same fraction of the collisional energy loss as for $\alpha_{s}^{fr}=0.5$.
Thus, for the reasonable bandwidths
in $m_{g}$ and $\alpha_{s}^{fr}$ the fraction of the collisional energy 
loss is small.

Note that due to the dominance of the radiative energy loss 
modeling the jet quenching with the collisional 
energy loss alone within the model of a particle undergoing Brownian motion
described by the Fokker-Planck equation
\cite{Mustafa-Thoma} does not make sense.
Evidently for an  accurate evaluation of the nuclear modification
factor the radiative and collisional energy losses 
must be treated on an even footing. One can expect a nontrivial interplay
of these two effects, say, it is clear that they cannot be additive, since
the collisional energy loss will suppress the effective in-medium gluon
formation length.  
The quantum nonlocal character of the induced gluon radiation 
makes the problem very complicated even at the level of the radiative 
energy loss. Presently the distribution in the induced energy loss which
is necessary for evaluation of the nuclear modification factor 
is usually calculated
assuming the independent gluon emission \cite{BDMS_RAA} which, however,
has not any serious theoretical justification.
Since the fraction of the collisional energy loss is small, it seems
reasonable to treat it as a perturbation. At qualitative level, neglecting the
nonadditivity, one can 
incorporate the collisional energy loss into this
model by a small renormalization of the QGP density according to the 
change in the $\Delta E$ due to the collisional energy loss.   
In \cite{Z04_RAA} we have 
described reasonably well the RHIC data on the nuclear modification
factor in $Au+Au$ collisions at $\sqrt{s}=200$ GeV by the induced gluon 
radiation alone with $\alpha_{s}^{fr}=0.7$. Inclusion of the 
collisional energy loss will require somewhat smaller value of 
$\alpha_{s}^{fr}$.
The results of this analysis will be presented elsewhere.

Note that although our approach does not treat accurately the region of soft 
momentum transfer $q\lsim m_{D}$ we can expect that this inaccuracy should be
small. Indeed, say, the relative contributions into collisional
energy loss of the soft region with $q\lsim 2\mu_{D}$ evaluated in our approach 
and in the HTL pQCD approach \cite{DE-Thoma} with accurate treatment of the
collective excitations are close. Also, even for low parton
energy $E\sim 5-10$ GeV the soft region gives
relatively small effect (about 30\% for $T\sim 250$ MeV). 
In any case, since the infrared effects should modify the dipole cross
section (\ref{eq:30}), which controls the induced radiation, 
and the probability of the collisional energy transfer 
approximately in the same way one can expect a good stability of the ratio
of the two mechanisms against the inaccuracy in the soft momentum region.

\vspace{.2cm}
\noindent {\bf 6}. 
In summary, we have performed the comparison of the radiative and 
collisional energy losses of energetic
quarks and gluons in an expanding quark-gluon plasma for 
RHIC and LHC energies.
The radiative energy loss has been calculated 
within the LCPI approach \cite{Z1}. To evaluate 
the collisional energy loss we have used the Bjorken model
of elastic binary collisions with an accurate
treatment of kinematics of the binary collisions.
The calculations have been performed with the same infrared cutoffs
and parametrization of the coupling constant for the radiative and collisional
energy loss, which is important for minimizing the theoretical 
uncertainties in the ratio of the radiative and collisional contributions.
Our numerical results demonstrate that for RHIC and LHC conditions
the fraction of the collisional energy loss is relatively small,
and decreases with energy. For gluons it is smaller than 
for quarks.

Our calculations show that the difference in the radiative 
energy loss for charm and light quarks is small. For this reason
the nuclear modification factor for light hadrons and $D$-mesons should be
approximately the same in the kinematical region where the light hadron 
spectra are dominated by the quark jets. At sufficiently large energies
the heavy quark energy loss becomes bigger than that for light quark.

\vspace {.7 cm}
\noindent
{\large\bf Acknowledgements}

\noindent
The author is grateful to the High Energy Group of the ICTP
for hospitality during his visit when this work was started.
This research is supported 
in part by the grants RFBR
06-02-16078-a and DFG 436RUS17/82/06.

\newpage

\begin{center}
{\Large \bf Figures}
\end{center}

\begin{figure}[h]
\begin{center}
\epsfig{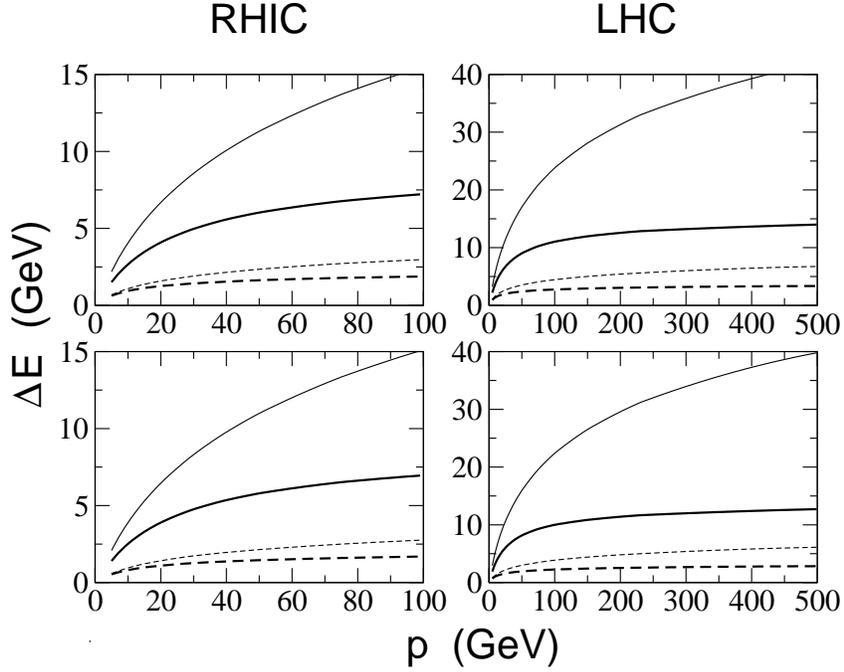}
\end{center}
\caption[.]
{
The light quark radiative (solid line) and collisional (dashed line) 
energy loss for RHIC (left) at $\sqrt{s}=200$ GeV and LHC (right) at 
$\sqrt{s}=5.5$ TeV conditions for $L=5$ fm.
The thick curves correspond to the running $\alpha_{s}$, and thin curves
to $\alpha_{s}=0.5$. The higher panels show the results for the 
$T$-independent Debye mass $\mu_{D}\approx 0.57$ GeV \cite{LH}, and the 
lower panels for the $T$-dependent Debye mass from the lattice calculations 
\cite{Bielefeld_Md}.
}
\end{figure}

\begin{figure}[h]
\begin{center}
\epsfig{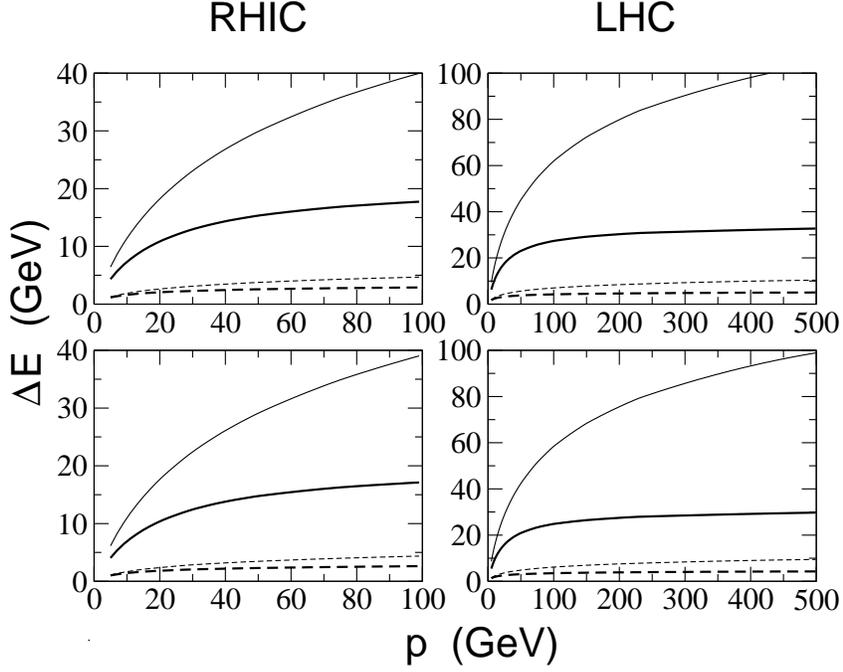}
\end{center}
\caption[.]
{
The same as in Fig.~1 but for gluon.
}
\end{figure}

\begin{figure}[t]
\begin{center}
\epsfig{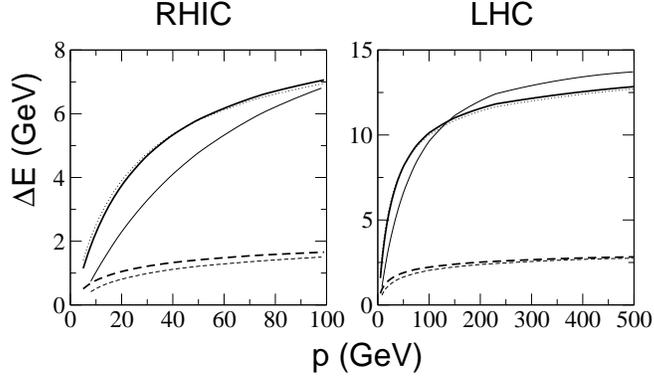}
\end{center}
\caption[.]{
The charm (thick curves) and bottom (thin curves) quark 
radiative (solid line) and collisional (dashed line) energy loss
for RHIC (left) at $\sqrt{s}=200$ GeV and LHC (right) at 
$\sqrt{s}=5.5$ TeV conditions for $L=5$ fm, $m_{c}=1.2$ GeV, 
$m_{b}=4.5$ GeV. The dotted line shows the radiative energy loss
for light quark.
The calculations were performed with the running $\alpha_{s}$ and 
the $T$-dependent Debye mass from 
\cite{Bielefeld_Md}.
}
\end{figure}

\end{document}